\documentclass[manuscript, nonacm]{acmart}
\AtBeginDocument{%
  }

\setcopyright{acmlicensed}
\copyrightyear{2026}
\acmYear{2026}
\acmDOI{XXXXXXX.XXXXXXX}

\acmISBN{978-1-4503-XXXX-X/2018/06}

\usepackage{svg}
\newtheorem{hyp}{Hypothesis}
\usepackage{booktabs}  
\usepackage{tabularx}
\usepackage{longtable}

\begin{document}

\title{Setting the Stage}
\subtitle{Enhancing Music Recommendations with Spoken Introductions to Stimulate Serendipity}

\author{Brett Binst}
\email{brett.binst@vub.be}
\orcid{0000-0002-0243-9952}
\affiliation{%
  \institution{imec-SMIT, Vrije Universiteit Brussel}
  \city{Brussel}
  \country{Belgium}
}

\begin{abstract}
People like to keep their music listening fresh and interesting, but recommender systems often struggle to support this need. 
Even when a recommended song is potentially refreshing and enriching, listeners may not \emph{perceive} it as such. 
Building on Silvia's appraisal theory of interest, this paper studies whether enhancing recommendations with spoken introductions --- immersive or informative --- can stimulate two distinct flavors of experienced serendipity: taste-broadening (sparking interest in music outside one's typical preferences) and taste-deepening (deepening interest in music within one's typical preferences).
Results from a large-scale user study show that both introduction types stimulate taste-broadening and taste-deepening serendipity.
Taste-broadening is most effectively stimulated through transportation (absorption in a narrative world), whereas taste-deepening is most effectively stimulated through cognitive elaboration (learning something new about the artist or the social context in which the music emerged).
We discuss how these findings could be translated into practice through curated taste-broadening playlists, where both introduction types appear promising, and taste-deepening playlists, where informative introductions seem particularly effective.
\end{abstract}



\maketitle

\section{Introduction}

\noindent\textbf{Imagine it's 1985: }You go to the record store in search of a new LP.
You browse through the covers when you suddenly encounter that cover that sparks your interest: a black woman in an Armani suit with a cigarette dangling from her red lips.
After waiting in line for the listening booth, you listen for just a few seconds, enough to know that you found your new favorite album.

\bigskip

\noindent\textbf{Fast-forward to 2005: }It's a Tuesday night and you're refreshing Bloglines instead of doing your homework.
\textup{\emph{Said the Gramophone}} has a new post praising this song called ``I'll Believe in Anything'' by a band you've never heard of, Wolf Parade.
You download the mp3, drag it onto your mp3 player.
You're not sure what to make of it at first.
But the next morning on the bus, you put it on again.
And again at lunch.
By the end of the week, it's the only thing you want to hear.

\bigskip

\noindent\textbf{Now it's 2026: }Ready to leave, you open Spotify and ask the DJ to play something energetic to start your day.
A familiar AI voice greets you by name, says it's pulling together an upbeat mix, and drops you into the soundtrack of your day.
The third song catches your attention.
It feels vaguely familiar, maybe you've heard it in a TikTok you scrolled past, or a YouTube video you saw.
You tap the $+$ to add it to your liked songs without breaking stride.
By Friday it's your most-played song of the month.

\bigskip
The emergence of the internet, and subsequently music streaming services, has fundamentally changed how we discover and experience music \cite{johansson_music_2025, ranaivoson_tuning_2023}.
Rather than purchasing physical albums or individual songs, many consumers today pay a subscription for access to a platform.
While physical formats, like vinyl, retain a charm for nostalgic or performative reasons \cite{whitehouse_taking_2023}, the dominant mode of listening is undeniably streaming \cite{mulligan_music_2025}.
The appeal is clear: streaming platforms offer the convenience of an essentially infinite music space, accessible anywhere as long as you have access to a laptop or smartphone, at a monthly rate lower than the price of buying an album \cite{jia_switching_2025}.
To help users navigate their large catalog, music streaming services integrate recommender systems on their platforms \citep{ricci_recommender_2022}.

However, recommender systems introduce their own challenges; a frequently discussed risk is that of trapping listeners in \emph{boredom bubbles} \citep{anderson_algorithmic_2020, shakespeare_reframing_2025}.
Indeed, research on music consumption stresses the importance of keeping the listening experience fresh \citep{greasley_keeping_2016}.
Therefore, literature on music recommendation stresses the importance of designing recommender systems for \emph{serendipity} \citep{kapoor_i_2015, li_understanding_2026, matt_escaping_2014, saito_fusion-based_2022, taramigkou_escape_2013, zhang_auralist_2012}.

Users experience serendipity when they fortuitously encounter content that is both refreshing and enriching \citep{binst_what_2025}.
While most researchers agree that it is important to design for serendipity, what serendipity exactly entails is less agreed upon \citep{kotkov_rethinking_2023, ziarani_serendipity_2021}.
\citet{binst_what_2025} address this issue by illustrating that serendipity expresses itself in different flavors.
In this paper, we will focus on two flavors that keep music listening fresh and engaging: taste-broadening and taste-deepening serendipity.

In the context of music, \emph{taste-broadening serendipity} entails fortuitously encountering a song that differs from a user's typical preferences but nevertheless sparks the listener's interest \citep{binst_what_2025}.
For example, while zapping through TV channels one evening, you stumble upon a movie about a musician's life that you normally do not listen to, but you find yourself genuinely intrigued by their music after learning their backstory.

\emph{Taste-deepening serendipity}, on the other hand, entails fortuitously encountering a song that aligns with the user's typical preferences but nevertheless is refreshing because the listener discovered the song from a fresh perspective, deepening the listener's interest \citep{binst_what_2025}.
For example, you stumble upon a TikTok that explains one of your favorite songs has a deeper meaning than you realized.

While accurately predicting and recommending interesting items is a necessary foundation for stimulating these serendipity flavors, it is not sufficient since it does not guarantee that the recommendations will be perceived as interesting.
Recommender systems must also convince the user that these recommendations are interesting in order to achieve taste-broadening or deepening.
A highly promising strategy for achieving this is through \emph{recommendation enhancements} \citep{behrooz_augmenting_2019, gabbolini_surveying_2024}.

Song recommendations can be enhanced in various ways, for example by adding contextual material such as audio or video.
However, it is a relatively underexplored area of research \citep{gabbolini_surveying_2024}.
For instance, it remains unclear which types of enhancement are effective in which contexts, and through which psychological mechanisms they operate.
Existing research appears to have examined only one type of music recommendation enhancement: the segue --- a short spoken text that links one song to the next \citep{gabbolini_generating_2021}.

This paper focuses on a different enhancement: spoken introductions.
Through a large-scale user experiment, three different types of introductions --- immersive, informative, and no introductions --- are tested to examine their effects on taste-broadening and taste-deepening serendipity and to explore the psychological mechanisms underlying these effects.
It extends prior work in which we studied the potential of these introductions in stimulating taste-broadening serendipity \citep{binst_let_2026}.
The present paper broadens the scope to include taste-deepening serendipity as well, enabling a direct comparison between the two flavors.

The results show that both immersive and informative introductions are effective in stimulating taste-broadening serendipity, but even more effective in stimulating taste-deepening serendipity.
Furthermore, the psychological mechanisms driving the two flavors differ: transportation is the strongest predictor of taste-broadening, while cognitive elaboration is the strongest predictor of taste-deepening.

\section{Related Literature}
\subsection{Theoretical Background: Music Psychology}

\citet{ward_same_2014} reveal a paradox: although most listeners state they desire novelty, when given the choice they decide to listen to the songs they already know.
In contrast, \citet{greasley_keeping_2016} observe that novelty also plays a central role in daily listening.
They find that listeners self-regulate their listening behavior to keep music fresh and avoid boredom bubbles \citep{greasley_keeping_2016}.

To understand this tension between exploration and exploitation in music listening, we must look at the cognitive mechanisms underlying it.
When listening to music, our brain constantly performs a process called \emph{predictive coding} through which we predict what will happen next in the song \citep{koelsch_predictive_2019}.
Successfully predicting what happens next results in a positive feeling \citep{green_listen_2012}.
Repeated listening to a song helps us in the process of predictive coding.
This leads to the \emph{mere exposure effect}, i.e., familiarity with music stimulates enjoyment \citep{green_listen_2012, peretz_exposure_1998}, which explains why people like to listen to the same old song as \citet{ward_same_2014} observed.

\begin{figure}[H]
  \centering
  \includegraphics[trim={1.91cm 7.58cm 12.18cm 4.70cm}, clip, width=0.45\columnwidth]{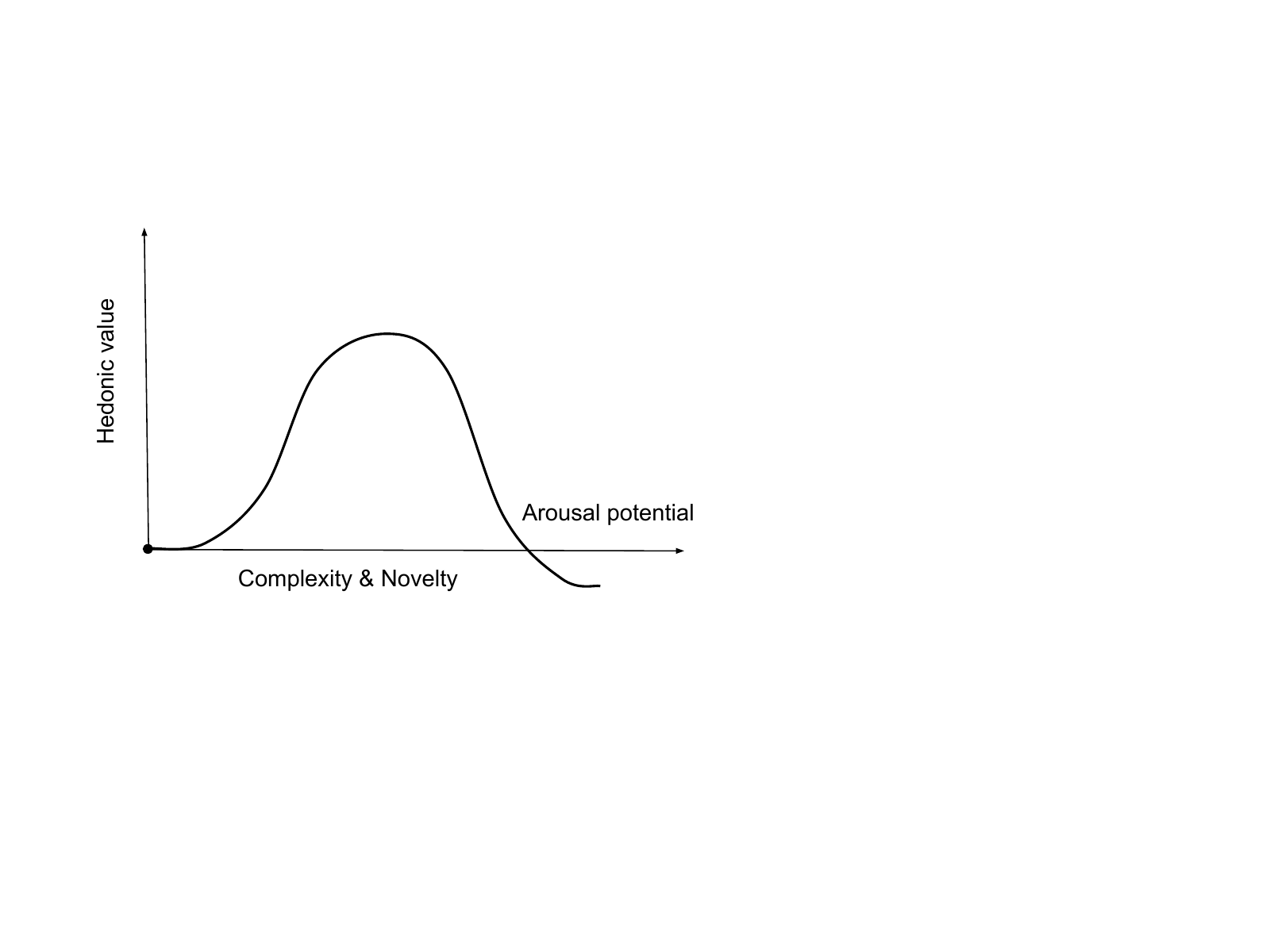}
  \caption{The relationship between Hedonic value and Complexity \& Novelty follows a Wundt-curve. Adapted from \citet{berlyne_novelty_1970}.}
  \label{fig:berlyne}
\end{figure}

However, the relationship between predictability and enjoyment is not linear, but follows an inverted-U shape \citep{green_listen_2012, koelsch_predictive_2019}.
This finding can be understood through \citeauthor{berlyne_novelty_1970}'s (\citeyear{berlyne_novelty_1970}) theory of aesthetic preference, which proposes that the hedonic value of a stimulus depends on its arousal potential, determined largely by complexity and novelty (see Figure~\ref{fig:berlyne}).
Familiarization shifts a stimulus toward the left-side of the curve \citep{berlyne_novelty_1970}.

The threshold of what is too predictable or too novel depends on one's cognitive capacity: the higher our cognitive capacity, the more open we are to new and complicated music; the lower our cognitive capacity, the more we crave predictability \citep{westgate_boring_2018, ward_same_2014}.
Hence, engagement with music requires a harmony between the cognitive demand of the music and the cognitive capacity of the listener \citep{westgate_boring_2018}.
When music becomes too familiar, and therefore too predictable, the cognitive demand will be low, ultimately resulting in boredom.
This explains why \citet{greasley_keeping_2016} observe that people take active precautions to keep music fresh.

\begin{figure}[H]
  \centering
  \includegraphics[trim={3cm 6cm 2cm 0cm}, clip, width=0.5\columnwidth]{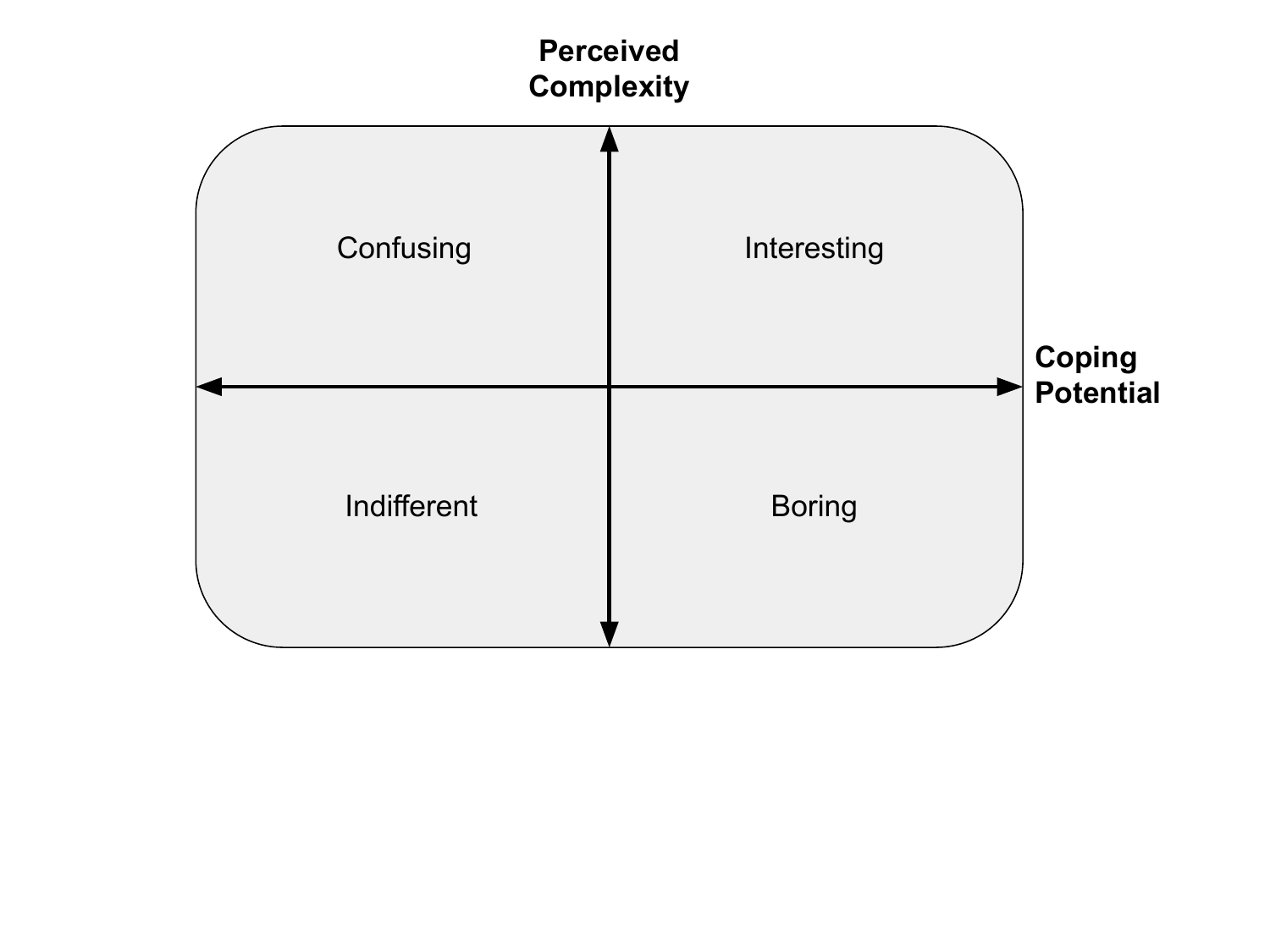}
  \caption{Schematic Representation of Silvia’s (2008) Appraisal Theory of Interest (source: Author)}
  \label{fig:silvia}
\end{figure}

Keeping music fresh is, at its core, a matter of stimulating interest \citep{hidi_four-phase_2006}.
Interest is a cognitive state that stimulates engagement and attention \citep{hidi_four-phase_2006}.
According to \citeauthor{silvia_interestcurious_2008}'s (\citeyear{silvia_interestcurious_2008}) appraisal theory, interest is triggered when a stimulus is perceived as ``complicated,'' yet within the observer's coping potential (see Figure~\ref{fig:silvia}).

Complicated is defined broadly here.
It encompasses stimuli that are sensorily rich (e.g., Beethoven's \emph{9th Symphony}), deep in meaning (e.g., Allen Ginsberg's \emph{Howl}), or counter-intuitive leading the observer to realize that things are more complicated than first believed (e.g., Robert Rauschenberg's \emph{Erased de Kooning Drawing}).
Importantly, these complicated stimuli will only be interesting if they are within the observer's coping potential, meaning observers feel capable of understanding them even if it might require a lot of effort \citep{silvia_interestcurious_2008}.

\citeauthor{silvia_interestcurious_2008}'s (\citeyear{silvia_interestcurious_2008}) theory implies that interest can be stimulated by manipulating perceived complexity and coping potential.
Research supports this: \citet{darda_impact_2023} find that providing contextual information about artworks in an exhibition increases both interest and liking, and \citet{silvia_what_2005} finds that providing background information about poetry sparks interest.
However, the most effective strategy likely depends on whether the music lies within or outside the listener's typical preferences.

\citeauthor{hidi_four-phase_2006}'s (\citeyear{hidi_four-phase_2006}) four-phase model distinguishes two types of interest: \emph{situational interest}, i.e., focused attention and an affective reaction triggered in the moment by an environmental stimulus, and \emph{individual interest}, i.e., a relatively enduring predisposition to reengage with particular content over time.
Situational interest aligns with becoming interested in music outside one's typical preferences, such as discovering a new genre in the car of a friend.
Individual interest aligns with interest in music within one's typical preferences, such as learning something previously unknown about one of your favorite bands.

These two forms of interest differ not only in their expression, but probably also in their related appraisals.
For music outside one's typical preferences, perceived complexity is likely high by default, so coping potential becomes the main lever: the listener needs guidance to make sense of what they are listening to.
For music within one's typical preferences, coping potential is likely already high, so what sustains and deepens interest is renewed complexity, such as novel angles, unexpected connections, or previously unnoticed details.

Serendipitous recommendations can stimulate either type of interest, and each gives rise to a distinct flavor of serendipity.
The next section describes these flavors.

\subsection{Taste-Broadening and Taste-Deepening Serendipity in Recommender Systems}

A user experiences serendipity when they fortuitously encounter content that is at the same time refreshing and enriching \citep{binst_what_2025}.
This experience can express itself in many flavors since the three core components --- fortuitous, refreshing, and enriching --- can express themselves in different ways.
This paper focuses on the two flavors that correspond to the types of interest described above: taste-broadening serendipity, which sparks a situational interest in music outside one's typical preferences, and taste-deepening serendipity, which deepens an individual interest in music within one's typical preferences.

Taste-broadening serendipity occurs when a user gets an unusual recommendation, i.e., something outside their typical preferences, that sparks their interest \citep{binst_what_2025}.
For example, imagine that Spotify's Discover Weekly recommends Bob Marley's \emph{Jamming} to you: reggae lies outside your usual listening habits, yet the song captivates you and becomes your favorite song of the week.
No individual interest has developed yet, but encountering the song has sparked a situational interest that may grow into one.

Taste-deepening serendipity, by contrast, occurs when a recommendation falls within the user's typical preferences yet reveals something refreshing and enriching.
For instance, imagine that you are a country fan, and your home screen surfaces Johnny Cash's 1968 live recording at Folsom Prison of \emph{Greystone Chapel} alongside a brief spoken introduction.
The introduction explains that the song was written by Glen Sherley, then a Folsom inmate, who was seated in the front row that morning with no knowledge that Cash --- having received a tape of the song only the night before --- was about to perform and immortalize it.
In this example, individual interest is already present; what the recommendation does is deepen it by uncovering something you were not yet aware of.

Although the distinction between taste-broadening and taste-deepening serendipity has only recently been articulated \citep{binst_what_2025}, previous research has studied music recommender systems designed for stimulating similar user experiences.
As the following sections show, several studies have examined how to stimulate taste-broadening.
Taste-deepening on the other hand is rarely studied.

\subsubsection{Designing for Taste-Broadening Serendipity}

A useful starting point into this literature is \citeauthor{liang_recommender_2019}'s (\citeyear{liang_recommender_2019}) distinction between pathway based and direct recommendation approaches for preference development. 

\paragraph{Pathway-based}

The first route consists of pathway-based systems, which try to create a smooth transition from a listener's current preferences to more distant musical territory.
For instance, \citet{taramigkou_escape_2013} created a system that broadens user preferences by computing the optimal trajectory from existing preference graphs to a chosen latent genre.
Similarly, \citet{flexer_playlist_2008} proposed a method for generating playlists that act as a seamless auditory bridge between a designated starting and ending song.
Building on this idea of step-by-step recommendation toward a target genre, \citet{nonnemaker_feedback-driven_2025} incorporated user feedback using Bayesian active learning, which further optimized and personalized the music exploration process.
The common thread among these studies is the assumption that music outside a user's typical preferences becomes more accessible when users are led toward it step by step.

\paragraph{Direct Recommendation Approaches}

In contrast, the second route consists of directly recommending a target genre in a way that increases users' interest.
We discuss four strategies.

The first strategy is through visualization and interactive design.
For instance, \citet{liang_interactive_2021} demonstrated that a contour plot --- which visually maps the spatial relationship between recommended tracks, the user's existing profile, and a novel target genre --- helps in giving users insight into new musical territories.
Closely related to visualization is the strategy of giving users interactive control over the recommender system.
Expanding on their visual models, \citet{liang_interactive_2021} observed that pairing a contour plot with an interactive mood slider effectively supports users in exploring the item space.
Providing this specific type of control directly leverages music's psychological function as a tool for mood regulation \citep{thoma_regulation_2006}.

A second strategy is context-aware recommendation, which assumes that music is more likely to be enjoyed when it fits the user's immediate listening context.
\citet{abdollahpouri_towards_2018} found that the preceding sequence of songs and advertisements heavily influences the success of algorithmic exploration.
Also the music recommender system by \citet{liang_interactive_2021}, discussed earlier, follows this logic.
By recommending songs based on one's mood, they should resonate better.
However, empirical research on context-driven exploration remains surprisingly limited \citep{aman_interacting_2017}.

A third strategy involves choice architecture and nudging.
Rather than changing the recommended items themselves, these approaches steer users toward exploratory behavior through the way options are presented.
\citet{liang_role_2021} showed that users are more likely to engage with distant genres when they are ranked at the top of a selection list, or when a genre-representativeness slider is defaulted to ``representative.''
However, follow-up work suggests that while such nudges can stimulate short-term exploration, their long-term effects remain modest \citep{liang_exploring_2022}.

A fourth strategy focuses on developing exploration-oriented algorithms diverging from the pathway approaches discussed earlier because they directly recommend a target genre.
\citet{zhang_auralist_2012} introduced an approach using Latent Dirichlet Allocation to model artist-associated listener communities.
By capturing listener diversity and applying a declustering algorithm, their system successfully recommends artists outside the user's established bubble.
Taking a multi-objective approach, \citet{mehrotra_algorithmic_2021} designed a system that balances the competing goals of discovery, familiarity, and similarity.
He demonstrated that applying hierarchical ordered weighted averaging to these goals outperformed baseline models, simultaneously boosting both discovery rates and overall user satisfaction \citep{mehrotra_algorithmic_2021}.

Although the design logic behind these recommender systems intuitively makes sense, they are not explicitly linked to theories about exploration.
In contrast, \citet{maccatrozzo_sirup_2017, maccatrozzo_role_2023} have applied Silvia's theory to the context of music recommendation.
They find that the cosine similarity calculated based on Linked Open Data paths can be used to calculate a novelty check, leading to more serendipitous recommendations \citep{maccatrozzo_sirup_2017}.
Furthermore, they find that diversity in the user profile can be used as a proxy for coping potential which is predictive for being more open to serendipitous recommendations \citep{maccatrozzo_sirup_2017}.
In this study, \citeauthor{silvia_interestcurious_2008}'s (\citeyear{silvia_interestcurious_2008}) appraisal theory is applied through a different lens; instead of using perceived complexity and coping potential as algorithmic checks for recommending items, the objective is to actively influence these perceptions through recommendation enhancements to stimulate interest as explained in Section~\ref{sec:rec-enhancements} and Section~\ref{sec:ce-tr-potential}.

\subsubsection{Designing for Taste-Deepening Serendipity}

The studies discussed above primarily focus on recommending genres outside the user's typical preferences, designed to stimulate taste-broadening.
In contrast, research specifically focused on taste-deepening --- encountering music that is refreshing and enriching, yet within the user's typical preferences --- is rare, and what exists has rarely been framed in those terms.
Two related design directions are nonetheless visible in the literature.

The first is the use of visualization tools to help users navigate the space within their typical preferences.
TastePaths \citep{petridis_tastepaths_2022} is a clear example: an interactive tool that renders a graph of related artists based on three artists the user frequently listens to, helping users discover hidden gems.
Although \citeauthor{liang_interactive_2021}'s (\citeyear{liang_interactive_2021}) contour-plot visualization was developed to stimulate taste-broadening, it shares the underlying logic: by visualizing the relationship between current preferences and recommended items, the navigation of the item space stimulates explorability \citep{bjorneborn_three_2017}, also within one's typical preferences.

A second design direction focuses on surfacing interesting songs in the long tail of the catalog --- niche items that fall within a user's typical preferences but receive little attention because of popularity bias \citep{celma_music_2010}.
While this literature is rarely framed in taste-deepening terms, the underlying logic is compatible: recommending a long-tail item within the user's typical preferences might stimulate taste-deepening.
In recommender system literature, serendipity is frequently operationalized precisely by measuring this dynamic: it is often calculated as the degree to which a specific system's recommendations deviate from a primitive, popularity-based baseline while remaining relevant to the user's taste \citep{de_gemmis_investigation_2015, karpus_serendipitous_2017, kotkov_survey_2016, maksai_predicting_2015, palumbo_entity2rec_2020}.

Researchers have proposed a range of algorithmic approaches to surface this kind of niche content.
One strategy is to leverage the behavior of users who are especially effective at discovering niche items themselves.
\citet{wang_serendipitous_2019} take this approach with innovator-based collaborative filtering in the e-commerce domain: a small minority of users --- the innovators --- naturally discover long-tail items on their own, and the system leverages their behavior as a signal for making recommendations.

A second strategy is to restructure sparse data so that tail items become recommendable.
\citet{park_long_2008} cluster tail items together so that their ratings can be aggregated, allowing the long tail to be recommended while individual-item collaborative filtering would lack sufficient signal.

A third strategy is to reformulate the relevance metric so that popularity is discounted.
\citet{yin_challenging_2012} treat the recommendation task as a random walk on a user-item graph.
Standard graph-based recommenders compute the walk from the user to candidate items, which gravitates toward popular items because they accumulate the most paths.
Computing the walk in the reverse direction --- from an item back to the user --- moves popularity into the denominator of the relevance score, promoting niche items connected to the user's taste.
A comprehensive review of long-tail recommendation is beyond the scope of this paper; for an in-depth treatment in the music domain, see \citet{celma_music_2010}.

Both visualization and long-tail recommendation primarily focus on helping the user discover new items.
However, this is not the only road towards taste-deepening; helping users uncover something new about music within their typical preferences can also spark a deeper interest.
In the next section, we discuss enhancing recommendations as a strategy more aligned with this uncovering condition of taste-deepening serendipity.

\subsubsection{Enhancing Recommendations}\label{sec:rec-enhancements}

Although the recommendation methods discussed above can potentially recommend items that carry the potential of broadening or deepening the user's taste, this does not necessarily imply that they will spark the user's interest.
Enhancing recommendations can help in highlighting the value of a recommended song. 
Furthermore, it has been argued to be effective in stimulating both taste-broadening \citep{behrooz_augmenting_2019} and taste-deepening serendipity \citep{gabbolini_play_2021}.
Enhancing song recommendations can be achieved in various ways, such as integrating videos into the listening experience (a feature recently adopted by platforms like Spotify) or providing background information about the tracks.

Enhancing song recommendations is an underexplored research domain \citep{gabbolini_surveying_2024}.
To date, the literature appears limited to a single strand of research on song recommendation enhancement: the usage of spoken \emph{segues} \citep{behrooz_augmenting_2019, gabbolini_user-centered_2022}.
Segues are short texts bridging consecutive songs, designed to make the listening experience more interesting \citep{behrooz_augmenting_2019}.
A whole playlist of songs connected by segues has been termed a \emph{music tour} \citep{gabbolini_user-centered_2022}.
\citet{gabbolini_user-centered_2022} find that these music tours are liked in the context of active listening.
They also find that good segues can spark interest in songs \citep{gabbolini_generating_2021}, and can help recommend songs in a fresh way \citep{gabbolini_play_2021}.
However, what constitutes a good segue and what psychological mechanisms underlie their effect remains underexplored.

This paper studies the psychological mechanisms underlying the effects of enhancing song recommendations.
However, instead of focusing on segues and the connections between tracks, the present study examines song introductions.
Introductions make it possible to study how to stimulate experienced serendipity while listening to a specific song, which segues cannot do.
Thus, this paper examines whether enhancing song recommendations with spoken introductions can stimulate taste-broadening and taste-deepening serendipity, and identifies the psychological mechanisms through which they operate.

\subsection{Transportation and Cognitive Elaboration as Pathways to Taste-Broadening and Taste-Deepening Serendipity}
\label{sec:ce-tr-potential}

Although \citeauthor{silvia_interestcurious_2008}'s (\citeyear{silvia_interestcurious_2008}) theory explains that interest is caused by appraising something as both complicated yet within one's coping potential, how to manipulate these two appraisals remains unclear.
This paper studies two potential pathways to influence these appraisals; through \emph{transportation} into the song's narrative and through \emph{cognitive elaboration}.

\subsubsection{Transportation}

Transportation refers to a state of deep absorption in a story \citep{green_narrative_2024}.
It consists of three main components: focused attention on the narrative, a sense of entering the world of the characters and scenery, and a temporary detachment from one's immediate surroundings and self-awareness \citep{van_laer_extended_2014}.

This state of transportation has several important effects.
First, transportation has a positive effect on processing fluency, that is, the ease with which content is processed and understood \citep{green_narrative_2024}.
Processing fluency, in turn, stimulates the positive evaluation of a stimulus \citep{schwarz_fluency_2018} and is associated with aesthetic pleasure \citep{reber_processing_2004}.
Transportation is also reciprocally related to meaningfulness: being transported can enhance the perceived meaningfulness of a narrative, while meaningful narratives can themselves stimulate transportation \citep{oliver_absorption_2017}.
Lastly, transportation reduces critical distance.
When people are transported into a narrative, they are generally less inclined to critically scrutinize its claims, which explains the power of stories as persuasive tools such as propaganda \citep{green_role_2000}.

Drawing on the effects outlined above, it is expected that stimulating transportation will enhance perceived meaningfulness and complexity.
Moreover, by improving processing fluency and providing a narrative structure that reduces ambiguity \citep{dahlstrom_using_2014}, transportation should also strengthen coping potential.
Transportation is hypothesized to be especially effective in stimulating taste-broadening serendipity, as it makes music outside the listener's typical preferences more accessible and lowers the psychological barriers to engagement.

There is some research on how to stimulate transportation.
Positive reviews have been shown to increase transportation \citep{isberner2019empowering, tiede2020reviews}, whereas cognitive load tends to reduce it \citep{green_narrative_2024}.
In addition, transportation is facilitated by narrativity, that is, the extent to which a text contains characteristics typical of stories, such as conflict, turning points, and unfolding events \citep{green_narrative_2024}.
However, if a narrative is perceived as corny and inauthentic, it is more likely to trigger affective resistance, i.e., an ``I'm not buying this story'' reaction, resulting in negative attitudes towards the `enhanced' content \citep{appel_affective_2022, appel_repelled_2019}.
Furthermore, when stories are believed to be written by GenAI, this also triggers affective resistance to the song, making transportation less likely \citep{messingschlager_creative_2024}.

Thus, existing research suggests that while transportation can be designed for by enhancing content with narratives, it is risky and can backfire if the narrative quality is poor.
The present research seeks to stimulate transportation using immersive introductions that focus the listener's attention on the narrative dimensions of recommended songs.
By foregrounding story-like elements and reducing interpretive ambiguity, such introductions may help transport listeners into the music.

\begin{hyp}
Immersive introductions stimulate taste-broadening and taste-deepening serendipity via a serial mediation path: increasing transportation, which subsequently increases both coping potential and perceived complexity.
\end{hyp}

\subsubsection{Cognitive Elaboration}

Enjoying music, especially less accessible genres like jazz and opera, requires the necessary background knowledge to understand the songs.
One possible way to make content more accessible is through cognitive elaboration, which can be seen as a targeted means of increasing cultural capital.
Cognitive elaboration is the active process of connecting new information to pre-existing broader schemas \citep{eveland_cognitive_2001}.

Cognitive elaboration impacts the aesthetic experience by facilitating \emph{cognitive mastering}.
According to \citet{leder_model_2004}, processing art or music can be framed as a problem-solving task where the listener actively attempts to find a satisfactory interpretation of the work.
When cognitive mastering is successful, the listener experiences a pleasurable sense of aesthetic insight \citep{leder_model_2004}.

Since cognitive elaboration is a learning process, it is logical to assume that it will stimulate listeners' coping potential with regard to the targeted song.
Furthermore, cognitive elaboration is expected to stimulate perceived complexity because it might reveal to listeners that there is more to the song than they initially realized.
This dynamic may be particularly relevant for music within the listener's typical preferences.
Therefore, cognitive elaboration is hypothesized to be especially effective in stimulating taste-deepening serendipity because it might uncover hidden meanings and details within the listener's typical preferences.

Several mechanisms help to stimulate cognitive elaboration.
One is personal relevance: when a song is framed in terms of universal themes or recognizably human experiences, it may become more relatable and personally meaningful, which in turn encourages deeper processing \citep{bartsch_making_2011, petty_elaboration_1986}.
Identification and empathy may play a similar role; learning about an artist's intentions or background story can encourage listeners to engage more deeply with the music \citep{igartua_identification_2016}.
Cognitive elaboration may also be triggered by creating a kind of ``cognitive itch,'' for example by drawing attention to a conflict, tension, or ambiguity in a song and subsequently helping the listener resolve it.
Finally, explicit elaboration prompts, such as reflective questions about the recommended content, have been shown to increase cognitive elaboration \citep{frauhammer_how_2025}.

These insights support the idea that cognitive elaboration can be deliberately designed for.
To this end, the present study aims to stimulate cognitive elaboration through informative introductions that provide background information, such as the inspiration behind the song, its social context, and hidden meanings within the lyrics.

\begin{hyp}
Informative introductions stimulate taste-broadening and taste-deepening serendipity via a serial mediation path: increasing cognitive elaboration, which subsequently increases both coping potential and perceived complexity.
\end{hyp}

The preceding sections discussed transportation and cognitive elaboration as distinct pathways for influencing perceived complexity and coping potential.
However, both psychological processes might also influence each other.
Indeed, research finds that transportation into the characters of short movies stimulates cognitive elaboration about the content \citep{igartua_film_2009}.
Therefore, this study also explores the correlation between these two psychological mechanisms, alongside the impact of both introduction types on them.
This results in the following model:

\begin{figure}[H]
  \centering
  \includegraphics[trim={0 5cm 0 0}, clip, width=0.644\textwidth]{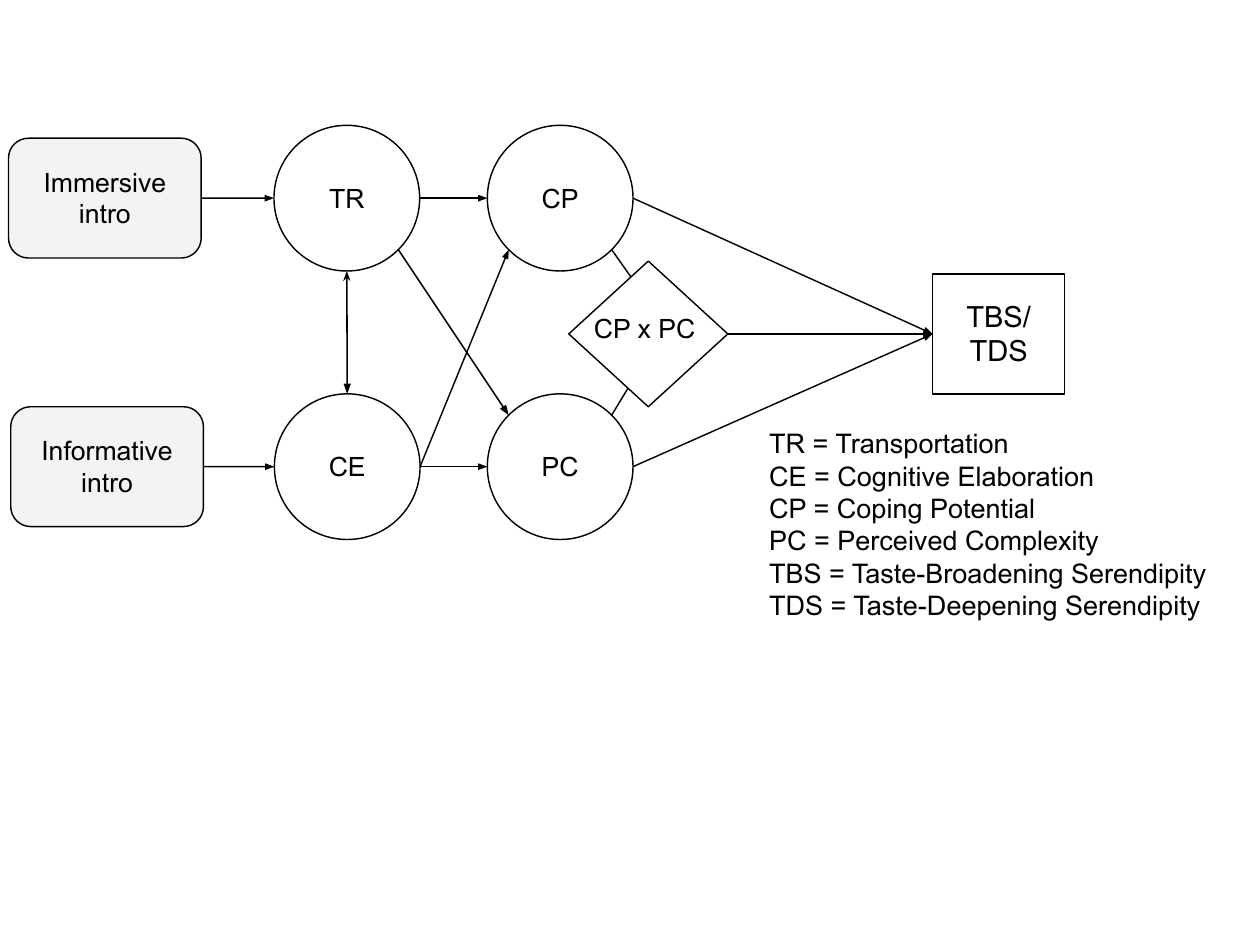}
  \caption{Path diagram illustrating the two pathways hypothesized to stimulate taste-broadening and taste-deepening serendipity.}
  \label{fig:hypotheses-model}
\end{figure}

\section{Methodology}
\subsection{Implementation}\label{sec:implementation}

The user study was done using a custom web-based application allowing remote, asynchronous participation across desktop and mobile platforms.
The application was hosted on Vercel (\url{vercel.com}), deploying a PostgreSQL database (\url{neon.com}) for GDPR-compliant data storage.
The system captured measures of user engagement (likes/dislikes, song skips, total listening time), and behavior (clickstream data and attention checks), as well as subjective measures of user experience.
The full code for the application can be found in \href{https://github.com/imec-smit-vub-mep-rec/Music-Introductions-Experimental-Environment}{this GitHub repository}\footnote{\url{https://github.com/imec-smit-vub-mep-rec/Music-Introductions-Experimental-Environment}}.

The impact of the different types of introductions on discovery was tested using a curated stimulus set of six genres, each containing three songs.
Genre selection was based on the four music-preference dimensions identified by \citet{rentfrow_re_2003} to ensure a diverse representation of different music preferences: reflective and complicated (Opera and Blues), intense and rebellious (Heavy Metal), upbeat and conventional (Country and Disco), and energetic and rhythmic (Reggae).

Each song received one of three introduction styles that I wrote:

\begin{itemize}
    \item \textbf{Informative}: providing background information, such as the inspiration behind the song, its social context, and hidden meanings within the lyrics.
    \item \textbf{Immersive}: foregrounding a possible narrative interpretation of the song by setting the scene.
    \item \textbf{Control}: no introduction.
\end{itemize}

These introductions were written to have a similar duration for the same song.
However, this was not strictly enforced since matching the length exactly would have meant sometimes lowering the quality of the introductions. 
As Table~\ref{tab:intro-duration} shows, on average, there is no systematic difference in duration between the two introduction types.

\begin{table}[H]
  \centering
  \caption{Mean duration (in seconds) of the introductions by genre and introduction type. Each genre contains three songs.}
  \label{tab:intro-duration}
  \begin{tabular}{lrr}
    \toprule
    Genre & Informative & Immersive \\
    \midrule
    Blues       & 30.2 & 39.8 \\
    Country     & 42.2 & 43.8 \\
    Disco       & 46.1 & 38.9 \\
    Heavy metal & 45.7 & 41.1 \\
    Opera       & 48.3 & 49.9 \\
    Reggae      & 35.9 & 29.6 \\
    \midrule
    All songs   & 41.4 & 40.5 \\
    \bottomrule
  \end{tabular}
\end{table}

I selected three songs per genre based on three criteria: how well they represent their genre, whether they tell a story that an immersive introduction can enhance, and whether verifiable background information about them is available for writing an informative introduction.

\begin{itemize}
    \item \textbf{Blues:} Hound Dog (Big Mama Thornton); I'd Rather Go Blind (Etta James); The Thrill Is Gone (B.B. King)
    \item \textbf{Country:} Folsom Prison Blues, Live (Johnny Cash); I'm So Lonesome I Could Cry (Hank Williams); Okie from Muskogee (Merle Haggard)
    \item \textbf{Disco:} Love to Love You Baby (Donna Summer); Smalltown Boy (Bronski Beat); YMCA (Village People)
    \item \textbf{Heavy metal:} Ace of Spades (Motörhead); Smoke on the Water (Deep Purple); War Pigs (Black Sabbath)
    \item \textbf{Opera:} Carmen, Act 1: Habanera (Bizet); La Bohème, Act 2: Quando Me'n Vò (Puccini); Porgy and Bess: Summertime (Gershwin)
    \item \textbf{Reggae:} Bam Bam (Sister Nancy); One Love / People Get Ready (Bob Marley \& The Wailers); Red Red Wine (UB40)
\end{itemize}

\begin{figure}[H]
  \centering
  \includegraphics[trim={0 3cm 0 0}, clip, width=\columnwidth]{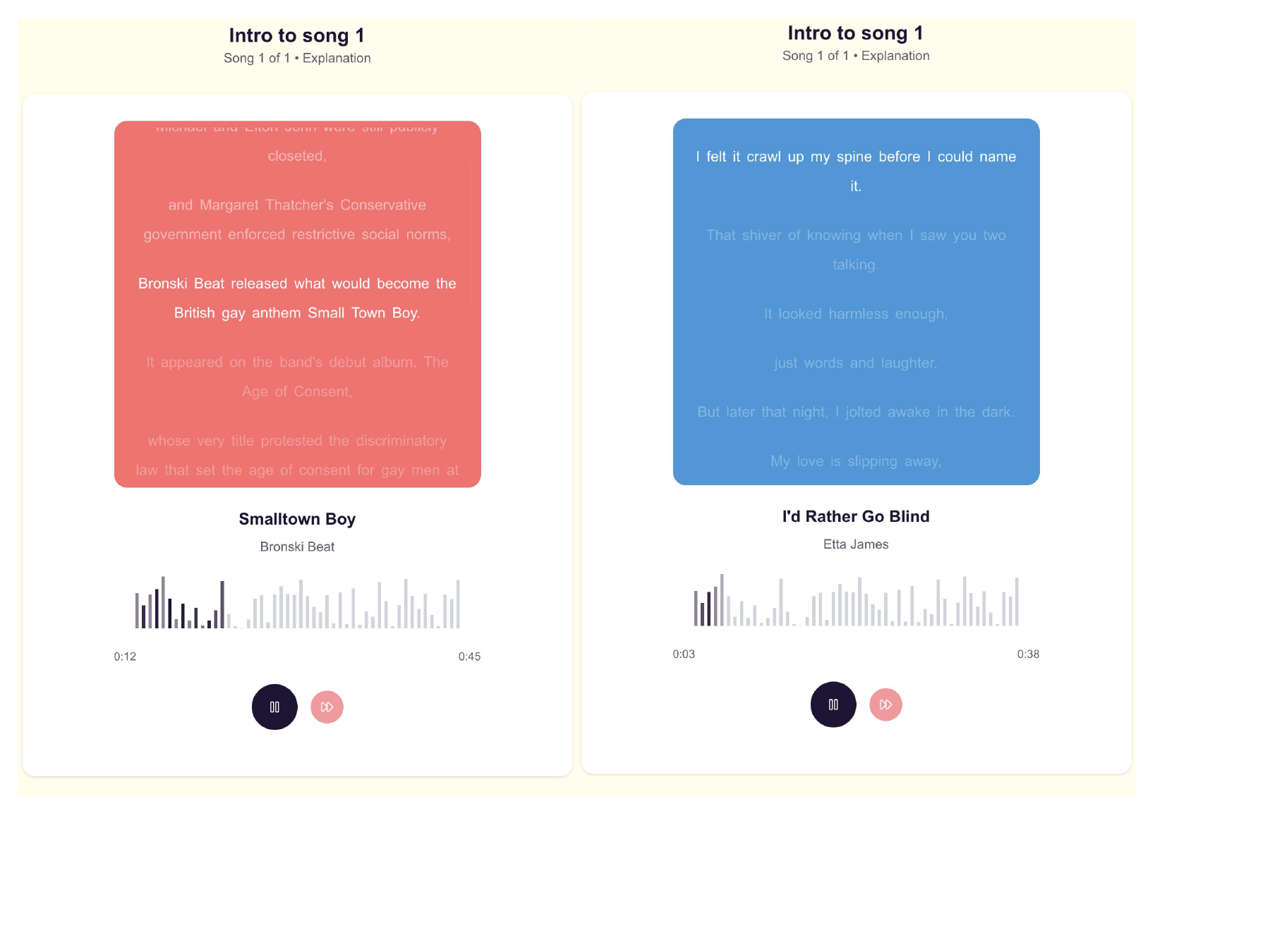}
  \caption{Left: An informative intro for ``Smalltown Boy''. Right: an immersive intro for ``I’d Rather Go Blind''}
  \label{fig:app}
\end{figure}

See Figure~\ref{fig:app} for an example of an informative and an immersive introduction used in the experiment.
These introductions were synthesized using Google Vertex AI Studio (Gemini 2.5 Text-To-Speech (TTS) model\footnote{\url{https://blog.google/innovation-and-ai/technology/developers-tools/gemini-2-5-text-to-speech/}}).
Finally, applying genre-specific vocal prompts and adjusting the speech rate to 1.25$\times$ resulted in a natural-sounding voice.
You can explore all the introductions as they were presented to the participants in \href{https://serendipity-experiment-beige.vercel.app/review}{this web app}.\footnote{\url{https://serendipity-experiment-beige.vercel.app/review}}

\subsection{Experimental Design}

Participants were randomly assigned to one of two conditions; either they were asked to select a genre they frequently listen to, or to select a genre they seldom listen to.
We refer to these as the \emph{within-preference} and \emph{outside-preference} conditions.
To account for the possibility that a participant might frequently listen to all or none of the six experimental genres, the assigned condition was validated at the song level through the statement, ``The song differs from the type of music I normally listen to,'' assessed via a 5-point Likert scale.

After choosing a genre, each participant listened to three songs from the genre followed by questions about their listening experience.
Each participant experienced all three introduction styles (control, informative, and immersive) but the order of the songs and introduction types was randomized; different sessions had a different order of songs, of introduction types, and therefore also a different combination of songs and introduction types.
This randomization effectively neutralizes systematic bias across conditions.
Furthermore, through multilevel statistical analyses and the within-subject design, individual differences, such as familiarity with the listened genre, were also controlled for.

The experimental procedure was structured into five phases:

\begin{enumerate}
    \item \textbf{Onboarding}: Upon informed consent, participants were asked to give some basic demographic information (country and date of birth).
    \item \textbf{Selecting a Genre}: Participants were randomly assigned to select either a genre they frequently listen to or one they seldom listen to.
    \item \textbf{Listening Task}: Participants listened to three distinct songs.
    While each participant listening to the same genre also heard the same three songs, both the order of the songs and their accompanying introduction style (informative, immersive, or no introduction) varied across sessions.
    Each of the three song trials consisted of an audio component followed by a survey.
    The audio player featured genre-specific visual cues, time-synced lyrics during spoken introductions, and an animated progress bar (see \autoref{fig:app}).
    \item \textbf{Post-Listening Measures}: Immediately after listening to each song, participants rated their experience on six dimensions (taste-broadening or taste-deepening serendipity, transportation, cognitive elaboration, coping potential, perceived complexity, quality of introduction) using 5-point Likert scales.
    \item \textbf{Post-Experiment Measures}: After completing all three song trials, participants filled in a final questionnaire measuring overall satisfaction and desire to listen to more songs from the explored genre.
    Finally, an open-text field asked participants to share their thoughts about the study; these responses are drawn upon in the discussion to help interpret the quantitative results.
\end{enumerate}

\subsection{Recruitment and Participants}\label{sec:recruitment}

The experiment was deployed in phases, beginning with a pilot on October 29, 2025, followed by a full release among our own network on November 20 (as a safety check for application stability before starting the Prolific deployment), and a Prolific deployment on December 2.
Data collection ended on December 8.

We recruited 475 participants, 402 via Prolific and 73 via snowball sampling.
To ensure a diverse sample on Prolific, we used stratified sampling to actively target diversity across education levels (1\% no formal education, 4\% secondary education, 17\% high school diploma, 13\% technical college, 40\% undergraduate degree, 21\% graduate degree, 4\% doctorate degree).
To obtain high data quality, we enforced strict exclusion criteria.
Five attention checks were embedded throughout the survey (e.g., ``Indicate `Strongly disagree' to continue'').
Participants who failed two or more attention checks were immediately redirected out of the study, and sessions with incomplete responses were excluded from the analysis.
In total, 23 participants were excluded (8 from Prolific, 15 from snowball sampling).
All numbers reported below refer to the final sample after exclusions.

The final sample ($N = 452$; Prolific: $n = 394$; snowball sampling: $n = 58$) was evenly split by gender, comprising 225 men, 221 women, and 6 individuals who identified as non-binary or other.
The median year of birth was 1994 (Q1~=~1982, Q3~=~1999).
The sample was geographically diverse, covering 39 different countries.
The largest cohorts originated from South Africa (17.0\%, $n = 77$), the United Kingdom (13.3\%, $n = 60$), and Belgium (9.5\%, $n = 43$).
After exclusions, assignment to the main conditions remained well balanced, with 51.5\% ($n = 233$) in the \textbf{within-preference} condition and 48.5\% ($n = 219$) in the \textbf{outside-preference} condition.
The distribution of participants across the six genre categories is detailed in Table~\ref{tab:genre_distribution}.

\begin{table} [H]
\centering \caption{Distribution of Participants Across Musical Genres After Exclusions ($N = 452$)} \label{tab:genre_distribution} \begin{tabular}{lccc} \hline \textbf{Genre} & \textbf{Absolute Frequency ($n$)} & \textbf{Relative Frequency} & \textbf{Percentage} \\ \hline Disco & 98 & 0.217 & 21.7\% \\ Blues & 89 & 0.197 & 19.7\% \\ Heavy Metal & 71 & 0.157 & 15.7\% \\ Reggae & 69 & 0.153 & 15.3\% \\ Country & 67 & 0.148 & 14.8\% \\ Opera & 58 & 0.128 & 12.8\% \\ \hline \end{tabular} \end{table}

\subsection{Measurement}\label{sec:measurement}
\autoref{tab:tab0} lists the scales presented to participants after each song and their corresponding reliability scores.
It also lists the items used to operationalize taste-broadening and taste-deepening serendipity.
All scale items are measured on 5-point Likert scales (strongly disagree to strongly agree).

Transportation was adapted from \citet{appel_transportation_2015} to fit the context of this study.
The coping potential and perceived complexity scales were developed for this study, starting from Silvia's conceptualization \cite{silvia_exploring_2006}.
The cognitive elaboration scale follows Eveland's conceptualization \cite{eveland_cognitive_2001}.
Piloting (see Section~\ref{sec:recruitment}) tested whether participants read the items as intended; those they misread or found confusing were refined or dropped.
All three scales have a coefficient $H > .90$, well above the .80 criterion for a well-defined, replicable latent variable \cite{rodriguez2016applying}.

For the taste-broadening and taste-deepening serendipity scales, internal reliability is not relevant since they are composite concepts.
Each composite combines the serendipity items listed in the lower block of \autoref{tab:tab0}, grouped by the dimension they capture: unusual and uncovering (as refreshing conditions), interesting (as enriching condition), and surprising and difficult to find (as fortuitous conditions).
A dimension is satisfied when the participant scores 4 or higher on its associated item, or on at least one of its items when it has several.
Following \citet{binst_what_2025}, taste-broadening serendipity was operationalized
\footnote{We also explored a fuzzy logic operationalization that maintained the same dimension thresholds ($>3$) but assigned weighted values to scores of 4 and 5 to create a continuous variable. This approach was discarded because the resulting distribution proved too difficult to model.}
as a binary variable that was one if the participant rated the listening experience as:

\begin{itemize}
    \item \textbf{Outside typical preferences} --- scoring 4 or higher on the unusual item.
    \item \textbf{Interesting} --- scoring 4 or higher on the interesting item.
    \item \textbf{Fortuitous} --- scoring 4 or higher on at least one fortuitous item.
\end{itemize}

All these conditions must be met, otherwise taste-broadening serendipity has value zero.\footnote{Note that this operationalization differs from \citet{binst_let_2026}, where taste-broadening serendipity was operationalized as the conjunction of unusual (i.e., outside typical preferences), interesting, and fortuitous \emph{or} uncovering items. In the present paper, uncovering is understood as part of the refreshing component of experienced serendipity and is therefore not included under the fortuitous criterion.}

Taste-deepening serendipity was operationalized, also following \citet{binst_what_2025}, as a binary variable that was one if the participant rated the listening experience as:

\begin{itemize}
    \item \textbf{Within typical preferences} --- scoring \emph{lower} than 4 on the unusual item.
    \item \textbf{Interesting} --- scoring 4 or higher on the interesting item.
    \item \textbf{Uncovering} --- scoring 4 or higher on at least one uncovering item.
    \item \textbf{Fortuitous} --- scoring 4 or higher on at least one fortuitous item.
\end{itemize}

All these conditions must be met, otherwise taste-deepening serendipity has value zero.

{\fontsize{8}{10}\selectfont
\renewcommand{\arraystretch}{0.95}
\setlength{\LTpre}{6pt}\setlength{\LTpost}{6pt}
\begin{longtable}{@{}p{0.88\linewidth}@{}r@{}}
    \caption{Scales Used to Operationalize Subjective User Perceptions and Experiences after Listening to a Song}
    \label{tab:tab0}\\
        \toprule
        \textbf{Scale and Items} & \textbf{Loading} \\
        \midrule
        \endfirsthead
        \multicolumn{2}{@{}l}{\textit{Table~\ref{tab:tab0} (continued)}} \\
        \toprule
        \textbf{Scale and Items} & \textbf{Loading} \\
        \midrule
        \endhead
        \bottomrule
        \endfoot
        \bottomrule
        \addlinespace
        \multicolumn{2}{@{}p{\linewidth}@{}}{\footnotesize \textit{Note:} After each item we communicate its standardized factor loading on the latent variable. $H$ represents the reliability coefficient for the scale. The serendipity items do not reflect a single latent variable. They are combined into the binary taste-broadening and taste-deepening serendipity composites described in the text, so no loadings or reliability coefficients are reported for them.} \\
        \endlastfoot

        \textbf{Coping Potential ($H = .95$)} & \\*
        I feel like I get the song & 0.893 \\
        The song makes sense to me & 0.913 \\
        I understand the song & 0.911 \\
        I feel like I get what the song is about & 0.893 \\
        I find the song confusing & 0.740 \\
        \addlinespace

        \textbf{Perceived Complexity ($H = .92$)} & \\*
        The song expresses a deep meaning & 0.662 \\
        This song feels superficial (without depth) to me & 0.778 \\
        There isn’t much depth to this song & 0.951 \\
        \addlinespace

        \textbf{Transportation} \cite{appel_transportation_2015} ($H = .86$) & \\*
        I could picture myself in the scenes of the song & 0.779 \\
        I was mentally involved in the song while listening & 0.815 \\
        The song affected me emotionally & 0.700 \\
        While listening to the song I had a vivid image of what it was about & 0.769 \\
        \addlinespace

        \textbf{Cognitive Elaboration ($H = .95$)} & \\*
        I gained new insights about the song or artist & 0.868 \\
        I learned something new about the song or artist I listened to & 0.954 \\
        I learned more about the background behind the song or artist & 0.912 \\

        \midrule
        \textbf{Serendipity Dimensions} \cite{binst_what_2025} & \\*
        \textit{Unusual} & \\*
        The song differs from the type of music I normally listen to & --- \\
        \addlinespace
        \textit{Interesting} & \\*
        It sparked my interest & --- \\
        \addlinespace
        \textit{Fortuitous} & \\*
        I felt surprised & --- \\
        I discovered something unexpected & --- \\
        I discovered something that might have been more difficult to discover without the help of this app & --- \\
        \addlinespace
        \textit{Uncovering} & \\*
        I experienced the song through a new perspective & --- \\
        I noticed details I didn't notice before & --- \\

\end{longtable}
}

\subsection{Analysis}
The effects of the introductions on taste-broadening and taste-deepening serendipity were estimated with multilevel Bayesian logistic regressions (Bernoulli likelihood, logit link), fitted in R with the brms package \citep{burkner_brms_2017} using the CmdStan backend via cmdstanr \citep{carpenter_stan_2017}.
Each flavor was analyzed on the ratings in which it can occur: taste-broadening serendipity on songs rated as outside the listener's typical preferences (a score of 4 or higher on the unusual item) and taste-deepening serendipity on songs rated as within them (a score lower than 4).
Introduction type was entered as a categorical predictor with no introduction as the reference category, and crossed random intercepts for session and song accounted for repeated ratings from the same session and of the same song.
The default priors of brms were used.
Each model was estimated with four chains of 2,000 iterations, the first 1,000 of which served as warm-up.

The indirect effects of the introductions via the two theoretical pathways illustrated in \autoref{fig:hypotheses-model} were estimated using multilevel Structural Equation Modeling (SEM).
SEM is a robust statistical framework to estimate latent constructs and all hypothesized paths in a single model, making it possible to explicitly model mediation of causal effects and to correct for measurement error of latent constructs \citep{knijnenburg_evaluating_2015}.
Lavaan in R was used to perform the SEM analyses \citep{rosseel2012lavaan}.
The Maximum Likelihood Robust (MLR) estimator was used because it is easy to interpret, provides robust standard errors, and takes into account clustering in data (in our case in individuals).
Our analysis results are available in \href{https://github.com/imec-smit-vub-mep-rec/setting_the_stage-analysis}{this GitHub repository}\footnote{\url{https://github.com/imec-smit-vub-mep-rec/setting_the_stage-analysis}}.

\section{Results}
\subsection{Introductions are Effective in Stimulating Taste-Broadening and Taste-Deepening Serendipity}

To evaluate the impact of the introductions on the user experience of listening to the introduced songs, a multilevel Bayesian regression was performed, controlling for clustering at both the song and session level.
The resulting odds ratios (OR) and 95\% credible intervals (CrI) are reported in Table~\ref{tab:serendipity_results}.

\begin{table}[htbp]
    \caption{Effect of Introductions on Taste-Broadening (TBS) and Taste-Deepening (TDS) Serendipity}
    \label{tab:serendipity_results}
    \begin{tabular}{llrrr}
        \toprule
        \textbf{Model} & \textbf{Term} & \textbf{OR} & \textbf{Lower CrI} & \textbf{Upper CrI} \\
        \midrule
        TBS & Immersive Introduction & 1.73 & 1.13 & 2.65 \\
        TBS & Informative Introduction & 2.61 & 1.70 & 4.07 \\
        TDS & Immersive Introduction & 3.23 & 1.87 & 5.83 \\
        TDS & Informative Introduction & 8.36 & 4.51 & 16.64 \\
        \bottomrule
    \end{tabular}
\end{table}

The results indicate that enhancing recommendations through introductions is an effective strategy for stimulating both flavors of serendipity.
When users were presented with songs from a genre outside their typical preferences, both types of introductions successfully stimulated taste-broadening serendipity.
However, the effect of the introductions on taste-deepening serendipity was even larger.
Compared to the no-introduction baseline, immersive introductions more than tripled the odds of experiencing taste-deepening serendipity (OR~=~3.23).
Informative introductions had an even larger effect, increasing these odds more than eightfold (OR~=~8.36).

\subsection{Coping Potential and Perceived Complexity Account for the Effects of Introductions on Taste-Broadening but not Taste-Deepening Serendipity}

\begin{figure}[H]
  \centering
  \includegraphics[width=0.900\textwidth]{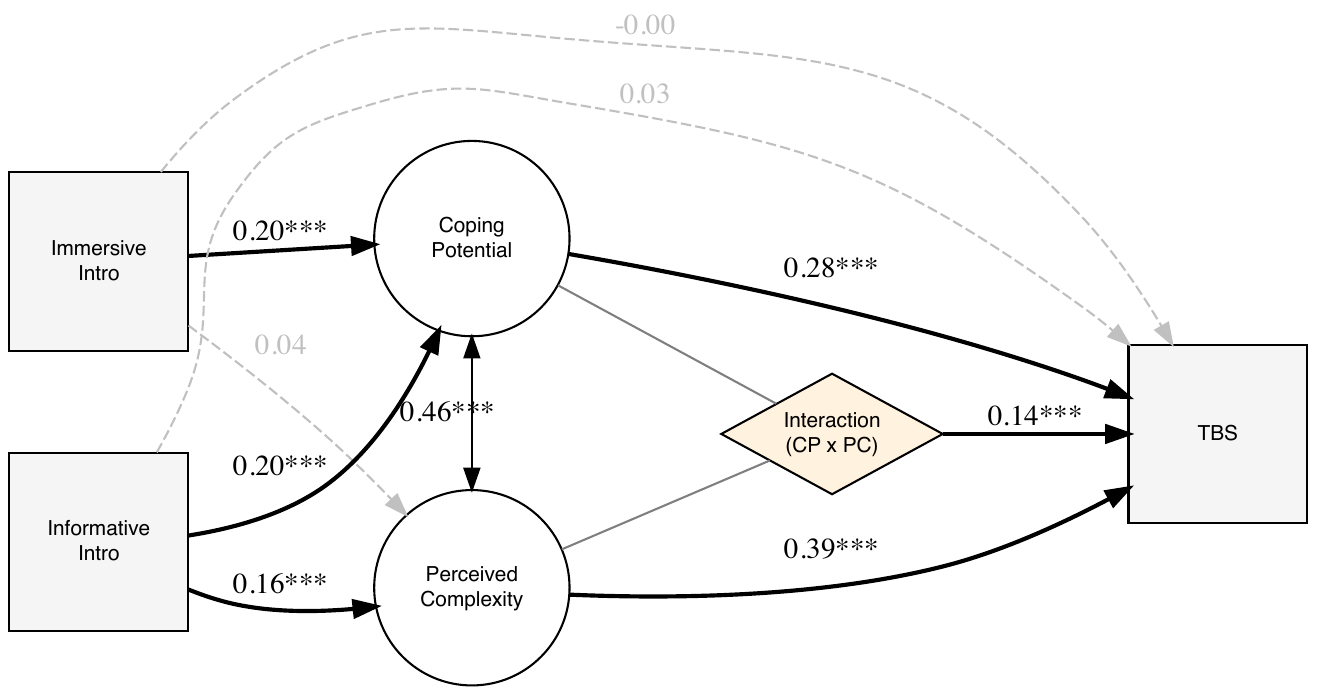}
  \caption{SEM path model predicting taste-broadening serendipity. Fit indices indicate good model fit: CFI~=~0.955, TLI~=~0.935, RMSEA~=~0.083, SRMR~=~0.036. Standardized coefficients are shown on each path; dashed gray paths are non-significant.}
  \Description{Structural equation model diagram showing two predictor boxes (Immersive Intro, Informative Intro) leading to two mediator circles (CP, PC) and their interaction. Both mediators and the interaction predict TBS. Standardized path coefficients are labeled on each arrow.}
  \label{fig:sem-tbs}
\end{figure}

To evaluate whether the effects of the introductions on taste-broadening and taste-deepening serendipity were accounted for by perceived complexity and coping potential, two separate path models were tested through multilevel SEM, one for the outside-preference condition (predicting taste-broadening serendipity), another for the within-preference condition (predicting taste-deepening serendipity).
The results shown in Figure~\ref{fig:sem-tbs} illustrate that the effects of immersive and informative introductions on taste-broadening serendipity are fully accounted for by indirect paths via coping potential and perceived complexity.
In line with \citeauthor{silvia_interestcurious_2008}'s (\citeyear{silvia_interestcurious_2008}) theory of interest, this pattern is consistent with the idea that the introductions stimulate taste-broadening by making the recommended songs seem both more understandable and richer in complexity.

\begin{figure}[H]
  \centering
  \includegraphics[width=\textwidth]{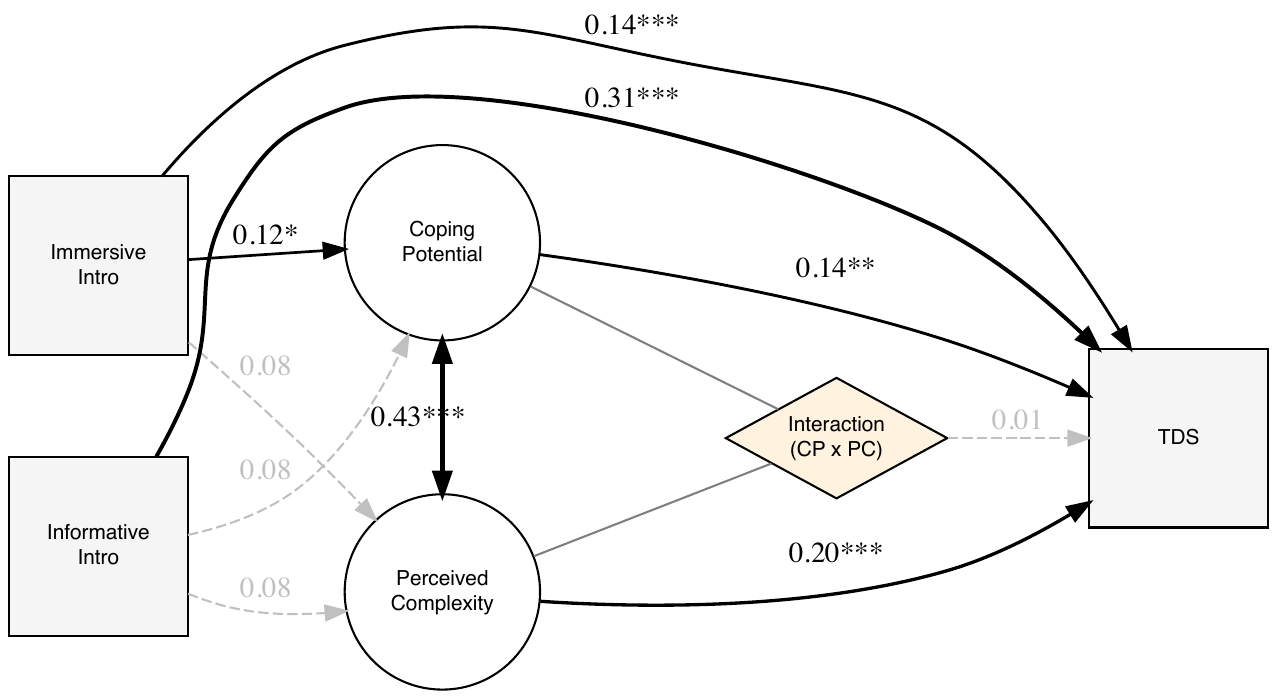}
  \caption{SEM path model predicting taste-deepening serendipity. Fit indices indicate good model fit: CFI~=~0.964, TLI~=~0.948, RMSEA~=~0.069, SRMR~=~0.042. Standardized coefficients are shown on each path; dashed gray paths are non-significant.}
  \Description{Structural equation model diagram showing two predictor boxes (Immersive Intro, Informative Intro) leading to two mediator circles (CP, PC) and their interaction, predicting TDS. Path coefficients are labeled on each arrow, with the direct paths from introductions to TDS remaining substantial.}
  \label{fig:sem-tds}
\end{figure}

However, the pattern differs for taste-deepening serendipity.
The results shown in Figure~\ref{fig:sem-tds} suggest that coping potential accounted for only part of the effect of immersive introductions, and that the strong effects of both immersive and, especially, informative introductions were not explained by coping potential and perceived complexity.

\begin{figure}[H]
  \centering
 \includegraphics[height=0.4\textheight, keepaspectratio]{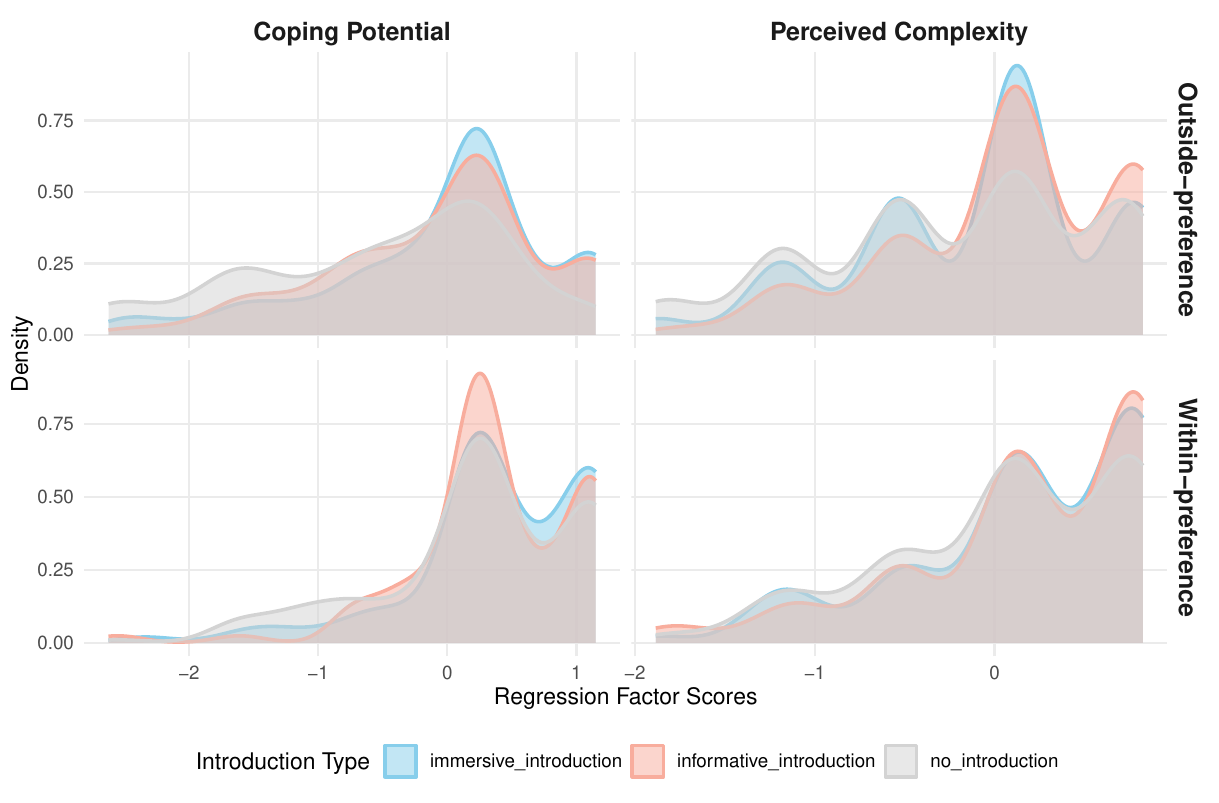}  
 \caption{Distributions of mediators by introduction type and condition.}
  \Description{Density plots of coping potential and perceived complexity in the outside-preference and within-preference conditions, with separate curves for each introduction type. In the within-preference condition, the distributions cluster toward the upper end of the scale even without an introduction.}
  \label{fig:density-pc-cp}
\end{figure}

One possible explanation is that songs from a genre within the listener's typical preferences are generally already perceived as highly complicated and comprehensible.
Comparing the distributions of these perceptions across the within-preference and outside-preference conditions (see Figure~\ref{fig:density-pc-cp}) indeed reveals evidence of a ceiling effect in the within-preference condition.
Even in the baseline no-introduction condition, these songs are associated with relatively high levels of coping potential and perceived complexity, leaving limited room for the introductions to further increase these perceptions.

\subsection{Transportation Drives Taste-Broadening, Cognitive Elaboration Drives Taste-Deepening}

To evaluate the underlying mechanisms driving serendipity, again two separate path models were estimated: one for predicting taste-broadening serendipity and another for predicting taste-deepening serendipity.

\begin{figure}[H]
  \centering
  \includegraphics[width=\textwidth]{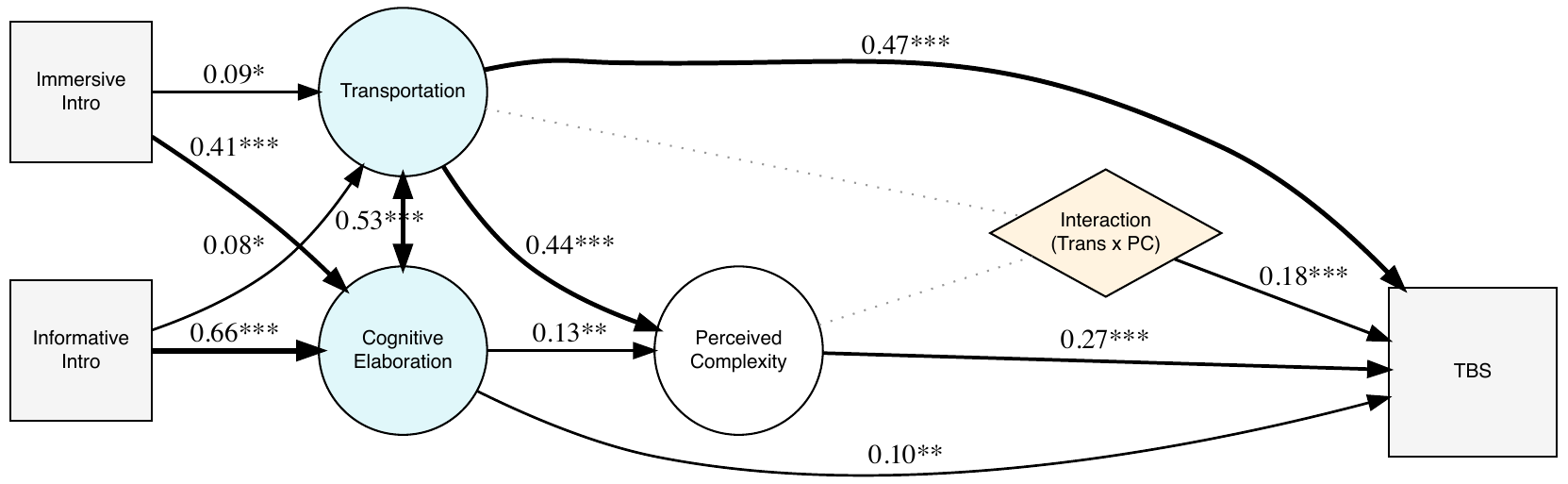}
  \caption{SEM path model predicting taste-broadening serendipity. Fit indices indicate good model fit: CFI~=~0.961, TLI~=~0.948, RMSEA~=~0.065, SRMR~=~0.060. Standardized coefficients are shown on each path; dashed gray paths are non-significant.}
  \Description{Path diagram with two predictor boxes (Immersive Intro, Informative Intro) feeding into transportation (TR) and cognitive elaboration (CE), which in turn predict perceived complexity (PC) and taste-broadening serendipity (TBS), with a covariance arrow between TR and CE.}
  \label{fig:sem-tbs-mechanisms}
\end{figure}

The path model predicting taste-broadening serendipity has a good overall fit (CFI~=~0.961, TLI~=~0.948, RMSEA~=~0.065, SRMR~=~0.060).
Coping potential was dropped from this model since including the direct effect of transportation makes coping potential redundant for predicting taste-broadening serendipity.
This indicates that transportation is the driving mechanism behind taste-broadening serendipity, and coping potential appears to be a by-product of high transportation, not a separate driver of taste-broadening serendipity.
Cognitive elaboration also has a significant direct effect on taste-broadening serendipity, but the effect-size is considerably smaller compared to transportation.
In addition, perceived complexity partially mediates the effects of both transportation and cognitive elaboration on taste-broadening serendipity.

An examination of the specific pathways (see Table~\ref{tab:unfamiliar_tbs}) reveals significant total effects for both immersive (std.\ effect~=~0.110, $p < .001$) and informative (std.\ effect~=~0.138, $p < .001$) introductions.

\begin{table}[htbp]
\caption{Pathway Effects on Taste-Broadening Serendipity for the Outside-Preference Condition}
\label{tab:unfamiliar_tbs}
\centering
\small
\setlength{\tabcolsep}{4pt}
\begin{tabular}{@{}llc@{}}
\hline
\textbf{Introduction} & \textbf{Pathway} & \textbf{Standardized} \\
\textbf{Type} & & \textbf{Effect} \\ \hline
\textbf{Immersive} & Introduction $\rightarrow$ Transportation $\rightarrow$ Taste-Broadening Serendipity & 0.044* \\
 & Introduction $\rightarrow$ Transportation $\rightarrow$ Perceived Complexity $\rightarrow$ Taste-Broadening Serendipity & 0.011* \\
 & Introduction $\rightarrow$ Cognitive Elaboration $\rightarrow$ Taste-Broadening Serendipity & 0.042** \\
 & Introduction $\rightarrow$ Cognitive Elaboration $\rightarrow$ Perceived Complexity $\rightarrow$ Taste-Broadening Serendipity & 0.014** \\
 & \textbf{Total Effect Immersive} & \textbf{0.110***} \\ \hline
\textbf{Informative} & Introduction $\rightarrow$ Transportation $\rightarrow$ Taste-Broadening Serendipity & 0.039* \\
 & Introduction $\rightarrow$ Transportation $\rightarrow$ Perceived Complexity $\rightarrow$ Taste-Broadening Serendipity & 0.010* \\
 & Introduction $\rightarrow$ Cognitive Elaboration $\rightarrow$ Taste-Broadening Serendipity & 0.066** \\
 & Introduction $\rightarrow$ Cognitive Elaboration $\rightarrow$ Perceived Complexity $\rightarrow$ Taste-Broadening Serendipity & 0.022** \\
 & \textbf{Total Effect Informative} & \textbf{0.138***} \\ \hline
\multicolumn{3}{l}{\small \textit{Note:} * $p < .05$, ** $p < .01$, *** $p < .001$}
\end{tabular}
\end{table}

\begin{figure}[H]
  \centering
  \includegraphics[width=\textwidth]{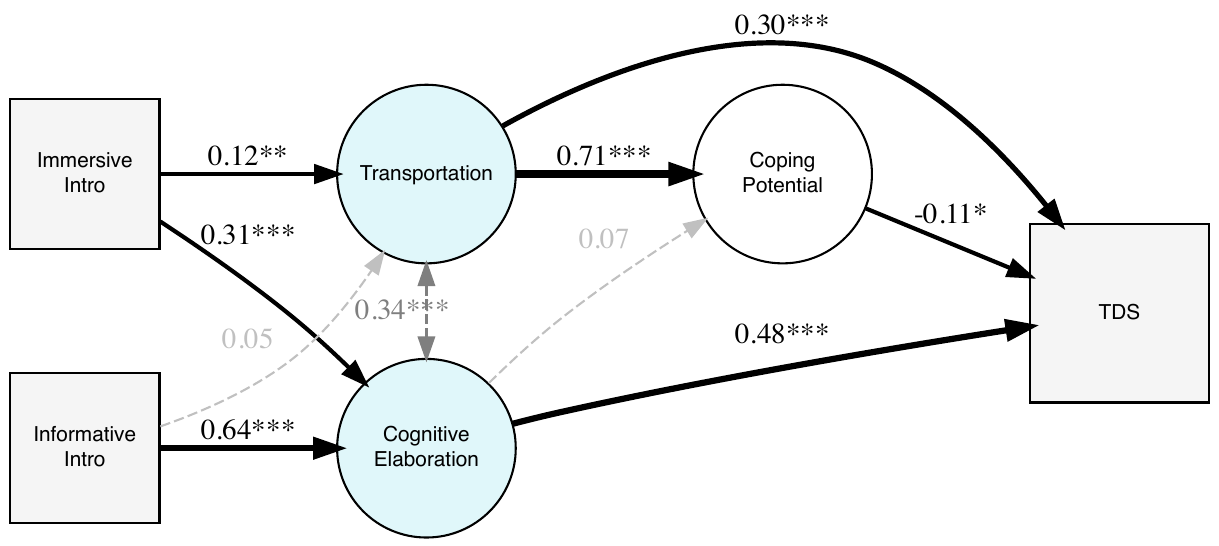}
  \caption{SEM path model predicting taste-deepening serendipity. Fit indices indicate excellent model fit: CFI~=~0.991, TLI~=~0.989, RMSEA~=~0.032, SRMR~=~0.024. Standardized coefficients are shown on each path; dashed gray paths are non-significant.}
  \Description{Path diagram with two predictor boxes (Immersive Intro, Informative Intro) feeding into transportation (TR) and cognitive elaboration (CE), which in turn predict coping potential (CP) and taste-deepening serendipity (TDS), with a covariance arrow between TR and CE.}
  \label{fig:sem-tds-mechanisms}
\end{figure}

The path model predicting taste-deepening serendipity shows excellent fit indices (CFI~=~0.991, TLI~=~0.989, RMSEA~=~0.032, SRMR~=~0.024).
Unlike the outside-preference condition, cognitive elaboration is the strongest predictor of taste-deepening serendipity in the within-preference condition.
However, transportation also has a strong direct effect on taste-deepening serendipity.
Surprisingly, there is also a negative  effect of coping potential on taste-deepening serendipity.

As detailed in Table~\ref{tab:familiar_tds}, the total effects are highly significant for both immersive (std.\ effect~=~0.171, $p < .001$) and informative (std.\ effect~=~0.311, $p < .001$) introductions.

\begin{table}[htbp]
\caption{Pathway Effects on Taste-Deepening Serendipity for the Within-Preference Condition}
\label{tab:familiar_tds}
\centering
\small
\setlength{\tabcolsep}{4pt}
\begin{tabular}{@{}llc@{}}
\hline
\textbf{Introduction} & \textbf{Pathway} & \textbf{Standardized} \\
\textbf{Type} & & \textbf{Effect} \\ \hline
\textbf{Immersive} & Introduction $\rightarrow$ Transportation $\rightarrow$ Taste-Deepening Serendipity & 0.037* \\
 & Introduction $\rightarrow$ Transportation $\rightarrow$ Coping Potential $\rightarrow$ Taste-Deepening Serendipity & $-0.009$ \\
 & Introduction $\rightarrow$ Cognitive Elaboration $\rightarrow$ Taste-Deepening Serendipity & 0.146*** \\
 & Introduction $\rightarrow$ Cognitive Elaboration $\rightarrow$ Coping Potential $\rightarrow$ Taste-Deepening Serendipity & $-0.002$ \\
 & \textbf{Total Effect Immersive} & \textbf{0.171***} \\ \hline
\textbf{Informative} & Introduction $\rightarrow$ Transportation $\rightarrow$ Taste-Deepening Serendipity & 0.016 \\
 & Introduction $\rightarrow$ Transportation $\rightarrow$ Coping Potential $\rightarrow$ Taste-Deepening Serendipity & $-0.004$ \\
 & Introduction $\rightarrow$ Cognitive Elaboration $\rightarrow$ Taste-Deepening Serendipity & 0.303*** \\
 & Introduction $\rightarrow$ Cognitive Elaboration $\rightarrow$ Coping Potential $\rightarrow$ Taste-Deepening Serendipity & $-0.004$ \\
 & \textbf{Total Effect Informative} & \textbf{0.311***} \\ \hline
\multicolumn{3}{l}{\small \textit{Note:} * $p < .05$, ** $p < .01$, *** $p < .001$}
\end{tabular}
\end{table}

Comparing the two models reveals that the effects of the introductions on transportation and cognitive elaboration are very similar, regardless of whether the listener is engaging with a genre within or outside their typical preferences.
In both cases, the introductions have a strong effect on cognitive elaboration while transportation proves more difficult to stimulate.
However, the primary mechanisms driving the two distinct types of serendipity clearly diverge.
In the within-preference condition, cognitive elaboration emerges as the strongest predictor of taste-deepening serendipity.
Conversely, in the outside-preference condition, transportation serves as the dominant driving force behind taste-broadening serendipity.

\section{Discussion}

A key challenge for music streaming platforms is preventing listeners from getting trapped in boredom bubbles \citep{anderson_algorithmic_2020}.
Indeed, music enthusiasts find it very important to keep their music listening fresh \citep{greasley_keeping_2016}.
Therefore, it is important to design music recommender systems for serendipity \citep{kapoor_i_2015, matt_escaping_2014, saito_fusion-based_2022, taramigkou_escape_2013, zhang_auralist_2012}.

Drawing on \citeauthor{silvia_interestcurious_2008}'s (\citeyear{silvia_interestcurious_2008}) appraisal theory of interest --- which identifies perceived complexity and coping potential as the two appraisals underpinning interest --- two types of spoken introduction were tested to influence these appraisals: immersive introductions, intended to stimulate transportation, and informative introductions, intended to stimulate cognitive elaboration.
Both types of introduction prove effective in stimulating taste-broadening and taste-deepening serendipity, and the results support the role of transportation and cognitive elaboration as the underlying psychological mechanisms.

These findings are discussed in more detail below.
The quantitative results are further enriched with open-text comments participants provided at the end of the experiment.

\subsection{Enhancing Recommendations through Song Introductions}

The introductions stimulate taste-broadening serendipity by introducing participants to genres outside their typical preferences.
As one participant mentioned: ``I never listen to heavy metal but this selection was better than I expected. I like the introductions which put us in the atmosphere of the song.''
Furthermore, fulfilling the intended design, the introductions motivated some participants to explore those genres further: ``opera is something I didn't think I would like but listening to the songs has changed my opinion and made me want to listen to more.''
Naturally, not all participants were persuaded: ``I HATE REGGAE''.

The introductions have a particularly strong effect on taste-deepening serendipity, providing a fresh listening experience for songs within participants' typical preferences, including songs they already knew, as the following comment illustrates: ``it was nice to listen when I know about the deep history behind the song\ldots small town boy hits different after that information.''
Several listeners mentioned that they really enjoyed experiencing their favorite music through a new perspective: ``That was fun, I enjoyed learning more about some of my favorite songs.''

While previous research hinted at the potential of song introductions to stimulate listener interest, the specific types of effective introductions and their underlying mechanisms have remained largely unexplored.
The current study addresses this gap by identifying two distinct psychological mechanisms that drive the efficacy of these introductions.

\subsection{Transportation Broadens Listeners' Taste}

As illustrated in Figure~\ref{fig:sem-tbs-mechanisms}, transportation was a particularly strong predictor of taste-broadening serendipity.
This effect can be attributed to the psychological mechanisms transportation triggers; it increases processing fluency and coping potential \citep{green_narrative_2024}, which in turn stimulates aesthetic pleasure \citep{reber_processing_2004, schwarz_fluency_2018} and elevates the perceived meaningfulness of the song \citep{oliver_absorption_2017}.

However, despite its efficacy, transportation is difficult to stimulate.
It relies on specific stimulus characteristics, such as narrativity \citep{green_narrative_2024} and authenticity \citep{appel_affective_2022}.
Furthermore, transportation requires a low cognitive load, yet, paradoxically, a literary style facilitates it \citep{green_narrative_2024}.
Given these stringent prerequisites, it is unsurprising that the immersive introductions yielded inconsistent effects, sometimes even backfiring.
Ultimately, listeners are most likely to experience taste-broadening serendipity when they are transported into unfamiliar music that they perceive as complicated or deep, a dynamic that aligns with the preference for authentic narratives \citep{appel_affective_2022}.

\subsection{Implications for Theory on Interest}

The findings suggest how \citeauthor{silvia_interestcurious_2008}'s (\citeyear{silvia_interestcurious_2008}) appraisal theory and \citeauthor{hidi_four-phase_2006}'s (\citeyear{hidi_four-phase_2006}) four-phase model of interest can be integrated.

For music outside the listener's typical preferences, taste-broadening serendipity requires sparking situational interest, and here \citeauthor{silvia_interestcurious_2008}'s appraisal theory fits the data well.
Perceived complexity and coping potential accounted for the effects of the introductions, which is consistent with the idea that introductions help listeners interpret the music they hear.
Transportation, however, was an even stronger predictor than coping potential.
This suggests that transportation does more than shape how listeners appraise a song: it enriches the listening experience itself, adding an aesthetic-affective component that a purely appraisal-based account of interest misses.

For music within the listener's typical preferences, taste-deepening serendipity requires deepening an existing individual interest, and appraisal theory has difficulty accounting for differences in experienced taste-deepening serendipity.
Listeners already perceived these songs as deep and comprehensible without an introduction, which left the introductions little room to stimulate either appraisal.
Instead, cognitive elaboration drives taste-deepening serendipity.
This suggests that what matters is whether an introduction uncovers a hidden layer in the song or lets listeners hear it from a fresh perspective.
Moreover, when controlling for transportation, coping potential was negatively associated with taste-deepening serendipity.
This is consistent with \citeauthor{berlyne_novelty_1970}'s (\citeyear{berlyne_novelty_1970}) inverted-U: for music within one's typical preferences, high coping potential may signal overfamiliarity rather than aid interpretation.

Together, these findings suggest that sparking situational and individual interest calls for different design goals: broadening listeners' taste benefits from helping them interpret the music, whereas deepening it benefits from surprising them with a fresh interpretation.

\subsection{Taste-Broadening and Taste-Deepening Playlists}

Participants voiced enthusiasm about integrating song introductions into music streaming platforms.
One participant, for instance, reflected: ``This was a genuinely fantastic experience and I appreciated knowing more about the context regarding songs, especially Ace of Spades. It definitely helped me enjoy the music way more and is something I would love to see implemented in apps.''
Spoken introductions could be deployed into music streaming as curated taste-broadening and taste-deepening playlists.

For taste-broadening playlists, featuring songs from genres or artists outside the listener's typical preferences, both informative and immersive introductions show promise.
For taste-deepening playlists, featuring songs from genres or artists within the listener's typical preferences, informative introductions appear to be particularly effective.

One participant suggested they would like to receive different introductions for the same song: ``That was fun, I enjoyed learning more about some of my favorite songs that evoke many memories for me from over the years. If this becomes an app it would be fun if every time you listened or chose to hear a fact there would be additional information so you could accumulate and learn more about the song from the composers and members of the groups.''

Despite the enthusiasm, previous research and feedback from our participants do indicate some critical points underpinning the successful implementation of spoken introductions.
As \citet{behrooz_augmenting_2019} state, voice characteristics are a major factor in the success of augmenting song recommendations with spoken introductions.
This is also reflected in the feedback on our experiment: ``the voice sounds like AI and not like a human; a natural human voice would be more engaging.''

This connects to a second concern: the negative perspective on AI.
\citet{messingschlager_creative_2024} observe affective resistance to stories when people believe them to be authored by GenAI.
This highlights the importance of authenticity as a requirement for introductions to be experienced as positive enhancements to recommendations.

A third point particularly concerns informative introductions.
Previous research suggests that facts are more interesting than opinions, especially if they are specific and non-trivial, but it is essential for users' trust that they are factually correct \citep{gabbolini_user-centered_2022}.

Lastly, prior work shows that augmenting song recommendations with spoken segues is valued in active rather than passive listening sessions \citep{behrooz_augmenting_2019}.
Recommending taste-broadening and taste-deepening playlists will likewise not always be valuable for the user, but mainly when they are actively listening to the music being played.

\subsection{Limitations and Future Directions}

\paragraph{Measurement}

The scales for coping potential, perceived complexity, and cognitive elaboration are self-developed, informed by \citet{silvia_exploring_2006} and \citet{eveland_cognitive_2001}.
The items were refined through extensive piloting to ensure they are understood as intended, and all three scales show high reliability (coefficient $H > .90$).
Nonetheless, as with any newly developed scale, further validation against related measures will strengthen confidence that they capture the intended constructs.
Furthermore, since the perceived complexity scale mainly focuses on the perceived depth of a song's meaning, it captures a concept that is narrower than what the label might suggest. 
It might have been more accurate to label the scale perceived depth. 
However, focusing on meaning is consistent with the coping potential items, which likewise focus on understanding what a song is about. 
Therefore, the findings reflect how listeners appraise a song's meaning.
Future research could distinguish semantic from structural complexity.

\paragraph{Ecological Validity}

To maximize internal validity and isolate the effects of the introductions, particular design choices were required.
Only a small but diverse stimulus set of three songs per genre was included, which made it possible to fully randomize the pairing of songs and introduction types and to write a high-quality introduction for each song.
Participants had to listen to the entire introduction, play at least 30 seconds of the song, complete all mandatory survey items, etc.
These requirements inevitably made the listening experience less naturalistic.
Future research could study the desire to listen to these introductions and their effect in a more realistic setting by, for example, A/B testing the introductions on an actual platform.

\paragraph{Stimulus Variability}

Given the small stimulus set, individual tracks may play a substantial role.
Nevertheless, because songs and introduction types are paired at random, they are independent and therefore individual songs cannot favor one introduction type over another.
However, the songs were deliberately selected because they afforded both informative and immersive introductions, for instance through an interesting backstory or a narrative that invites transportation.
This might limit generalization since other songs may lend themselves less well to such enrichment. 
The effectiveness of an introduction type likely also depends on how the introduction was written for a particular song.
Future research could test introductions on a larger and more varied set of songs, including songs with less narrative potential, and examine which characteristics of songs and introductions drive interest, perceived complexity, transportation, and cognitive elaboration.

\paragraph{Long-Term Impact}

Only the intention to listen to more songs from the same genre was measured, which is a proxy of actual taste-broadening and taste-deepening.
A longitudinal design could better capture whether the recommendations are truly enriching by stimulating increased interest and engagement in the long-term.

\paragraph{Variation in Effects}

Another promising direction for future research is to examine how personal characteristics shape the effectiveness of introductions in stimulating taste-broadening and taste-deepening serendipity.
Several traits have been linked to transportation, including imaginative engagement \citep{black_imaginative_2022}, transportability \citep{green_need_2020}, openness to experience \citep{silvia_personality_2011}, need for cognition \citep{appel_transportation_2010, thompson_sometimes_2012}, and need for affect \citep{maio_need_2001, thompson_sometimes_2012}.
Less is known about the traits that influence cognitive elaboration, except that a high need for cognition stimulates it \citep{cacioppo_central_1986}.
Whether and how these traits moderate the effects of different introduction types remains, to date, an open question.

\paragraph{Large-Scale Implementation}

Although human-authored introductions formulated by passionate music experts likely produce the best effects, this manual approach limits scalability, especially if users desire multiple distinct introductions per song.
Future research could explore automating this generation process using Large Language Models integrated with knowledge bases.
For instance, literature, documentaries, and music magazines offer a wealth of surprising information.
Furthermore, integrating a Text-to-Speech architecture, as described in Section~\ref{sec:implementation}, could facilitate the scalable deployment of enhancing recommendations with spoken introductions.

\section{Conclusion}

A key challenge for music recommendation is keeping the listening experience engaging and helping listeners escape boredom bubbles.
Even though the recommended music might be interesting to users, this does not automatically mean that these songs will spark their interest.
Therefore, a large-scale user experiment was conducted to test whether enhancing song recommendations with either informative, immersive, or no introduction stimulates a serendipitous listening experience.
Results indicate that transporting listeners into songs outside their typical preferences is the most effective way to broaden their taste, while cognitive elaboration with songs within their typical preferences is the most effective way to deepen listeners' taste.
Participants voiced their enthusiasm for implementing introductions on music streaming platforms, possibly through taste-broadening and taste-deepening playlists.

\section{Ethical and Human Subjects Considerations}
The study protocol was reviewed and approved by the Ethics Committee of Vrije Universiteit Brussel. 
Informed consent was obtained from all participants before starting the experiment. 
Participants were explicitly informed of their right to withdraw from the study at any time without penalty.


\begin{acks}
This work is supported by the Research Foundation Flanders (FWO) under grant numbers S006323N and V450225N.
I am grateful to all the participants for generously contributing their time and insights to this study. 
Furthermore, I thank the members of the reading group at SMIT for their thorough reviews of the manuscript and the many constructive suggestions.
Special thanks go to Ulysse Maes, Annelien Smets, Martijn Willemsen, and Olivia Willems, whose contributions were essential to this study.
\end{acks}

\bibliographystyle{ACM-Reference-Format}
\bibliography{library}

\appendix

\end{document}